\documentclass[aps,prl,twocolumn,tightenlines,floatfix,showpacs,superscriptaddress]{revtex4-1}
\usepackage{amsmath}
\usepackage{amsfonts}
\usepackage{amssymb}
\usepackage{graphicx}

\begin{document}

\title{Critical branching processes in digital memcomputing machines}

\author{Sean R. B. Bearden}
\affiliation{Department of Physics, University of California, San Diego, La Jolla, CA 92093}

\author{Forrest C. Sheldon}
\affiliation{Department of Physics, University of California, San Diego, La Jolla, CA 92093}

\author{Massimiliano Di Ventra}
\email{email: diventra@physics.ucsd.edu}
\affiliation{Department of Physics, University of California, San Diego, La Jolla, CA 92093}

\begin{abstract}
Memcomputing is a novel computing paradigm that employs time non-locality (memory) to solve combinatorial optimization 
problems. It can be realized in practice by means of non-linear dynamical systems whose point attractors represent the solutions of the original problem. 
It has been previously shown that during the solution search digital memcomputing machines go through a transient phase of avalanches (instantons) that promote dynamical long-range order.  By employing mean-field arguments we predict that the distribution of the avalanche sizes follows a Borel distribution typical of critical branching processes with exponent $\tau= 3/2$. We corroborate this analysis by solving various random 3-SAT instances of the Boolean satisfiability problem. The numerical results indicate a power-law distribution with exponent $\tau = 1.51 \pm 0.02$, in very good agreement with the mean-field analysis. This indicates that memcomputing machines self-tune to a critical state in which avalanches are characterized by a branching process, and that this state persists across the majority of their evolution.

\end{abstract}

\maketitle

\section{Introduction}

Unconventional computing paradigms that employ physical properties to compute specific problems are emerging as an important research direction 
in Physics \cite{nielsen_chuang_2010,book_neuromorphic,review_neuromorphic}. One such paradigm is 
{\it memcomputing}~\cite{diventra13a,DMMperspective}, which employs time non-locality (memory) to both process and store information on the same physical 
location. The digital version of this paradigm has been introduced to specifically tackle combinatorial optimization problems~\cite{DMM2}. Digital 
memcomputing machines (DMMs) can be physically realized as non-linear dynamical systems whose point attractors represent the solutions of the problem to be solved. 

Since DMMs are non-quantum systems, their equations of motion can be efficiently integrated numerically. Results from these simulations have already 
demonstrated that DMMs perform orders of magnitude faster than traditional algorithmic approaches on a wide variety of combinatorial optimization problems~\cite{exponential2017speedup,DMMperspective,AcceleratingDL,ILP,spinglass}. 

Subsequently, by employing topological field theory~\cite{topo}, it was  shown that 
the physical reason behind this efficiency rests on the dynamical long-range order that develops during the transient dynamics where avalanches (instantons 
in the field theory language) of different sizes are generated until the system reaches an attractor~\cite{spinglass}. The transient phase of the solution search of DMMs therefore resembles that of several phenomena in Nature, such as earthquakes \cite{earthquakes}, solar flares \cite{solarflares}, quenches~\cite{pruessner_2012}, etc. Since 
all these phenomena show scale-free properties in the probability distribution of the avalanche sizes, it is natural to ask whether DMMs would also share this 
property. 
In this paper, we indeed show that  
the transient dynamics of a DMM  are characterized by a {\it critical branching process}. We first provide a general mean-field analysis to argue that the probability 
distribution of the avalanche sizes should be a critical Borel distribution with exponent $\tau=3/2$~\cite{borelApprox3}, irrespective of the problem to solve, 
namely it is an intrinsic feature of DMMs. We then support these results with 
numerical simulations of DMMs' equations of motion applied to the solution of Boolean satisfiability (SAT) instances. We have chosen to work with randomly-generated, satisfiable 3-SAT benchmark instances precisely to ensure that any feature produced by our analysis is a feature  of the dynamics of DMMs, rather than a feature of the SAT instances solved. 

Random 3-SAT belongs to the class of propositional logic in which a formula of Boolean variables must hold true for the problem to be satisfiable \cite{bradley2007digital}.
Propositional variables appear as literals in the formula, where a literal is a variable or its negation.
Satisfiability problems are traditionally represented in ``conjunctive normal form'' (CNF), i.e., a conjunction (AND) of disjunctions (OR) of literals~\cite{computational_complexity_book}. 
A disjunction of literals is referred to as a clause. Therefore, a 3-SAT problem is one in which all clauses contain three distinct literals, of which none is a negation of the others. 

\begin{figure}[t!]
	\centering
	\includegraphics[width=8.5cm]{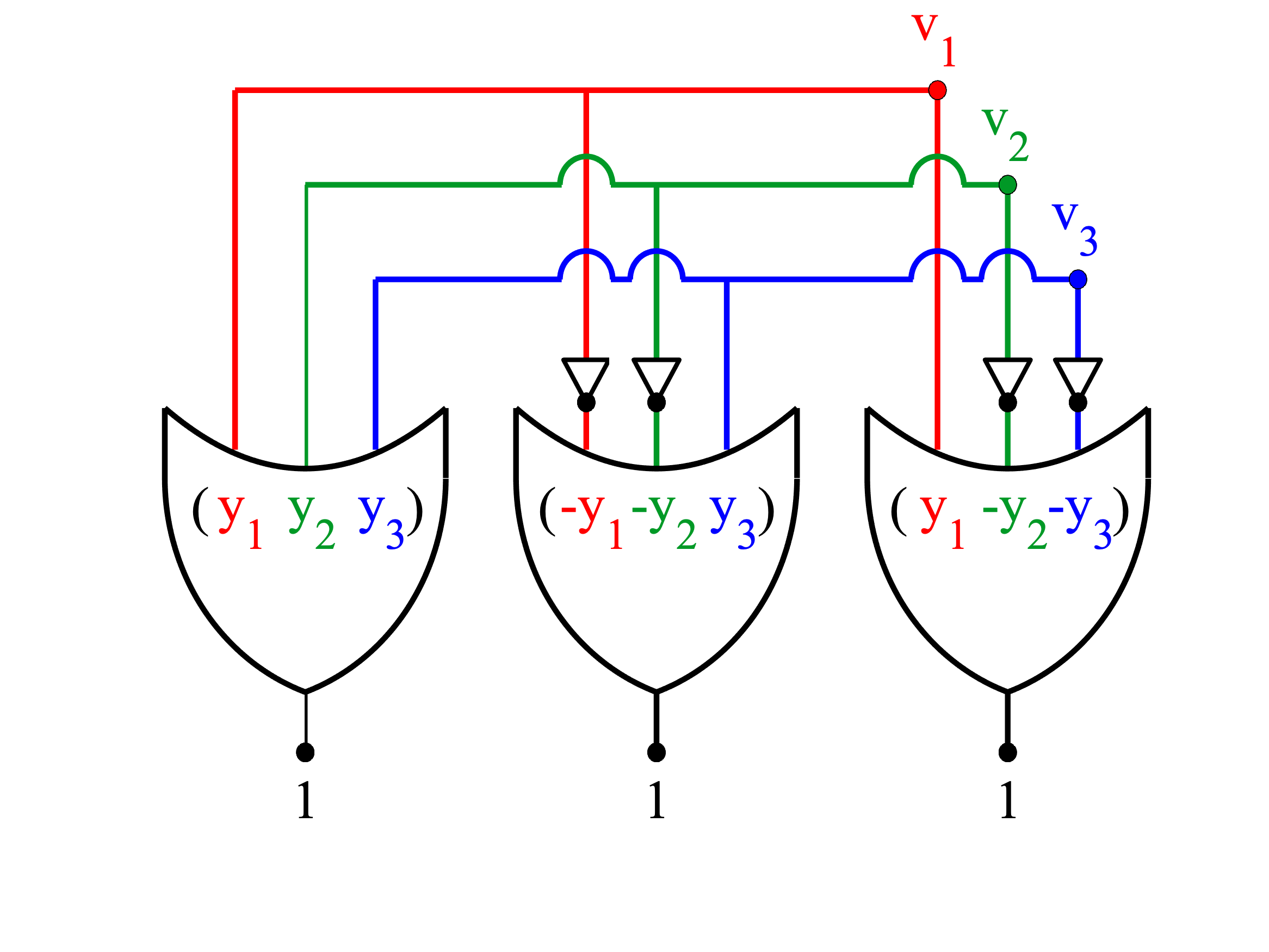}
	\caption{Example of a Boolean circuit, in conjunctive normal form (CNF), representing a 3-SAT. The three OR clauses (seen inside the gates) are then converted to self-organizing logic gates (SOLGs) where the propositional variables $y_i$ are represented as electrical voltages $v_i$. If a literal is the negation of a variable, then the associated ``input'' terminal on that gate must pass through a NOT gate (triangle symbol) before the terminal is connected to other terminals sharing the same variable. The traditional output of the SOLG-OR is forced to be true (logical 1), because all clauses must be true to satisfy a Boolean proposition in CNF. }
	\label{fig:circuit}
\end{figure}

A CNF formula has a simple Boolean circuit representation~\cite{computational_complexity_book}. An example for a 3-SAT with three clauses is reported in Fig.~\ref{fig:circuit}. A DMM that solves the 3-SAT, say, of Fig.~\ref{fig:circuit} can then be realized as an electrical circuit with memory (see Eqs.~(\ref{eq:voltages}),~(\ref{eq:xfast}), and~(\ref{eq:xslow}) below) where each 
variable of the 3-SAT problem is represented by a voltage (we represent with $+1$ the logical 1 and with $-1$ the logical 0, in arbitrary units), and each traditional OR gate of Fig.~\ref{fig:circuit} is replaced by a ``self-organizing'' OR gate~\cite{DMM2}, namely one that always attempts to dynamically satisfy the logical OR truth table at its terminals. Since the problems we are seeking to solve are satisfiable, the ``out'' terminals of the CNF formula in Fig.~\ref{fig:circuit} 
are set to be logically true, hence the voltages at those terminals are kept fixed at +1. 

\section{Mean-field analysis}

\begin{figure}[t!]
	\centering
	\includegraphics[width=8.5cm]{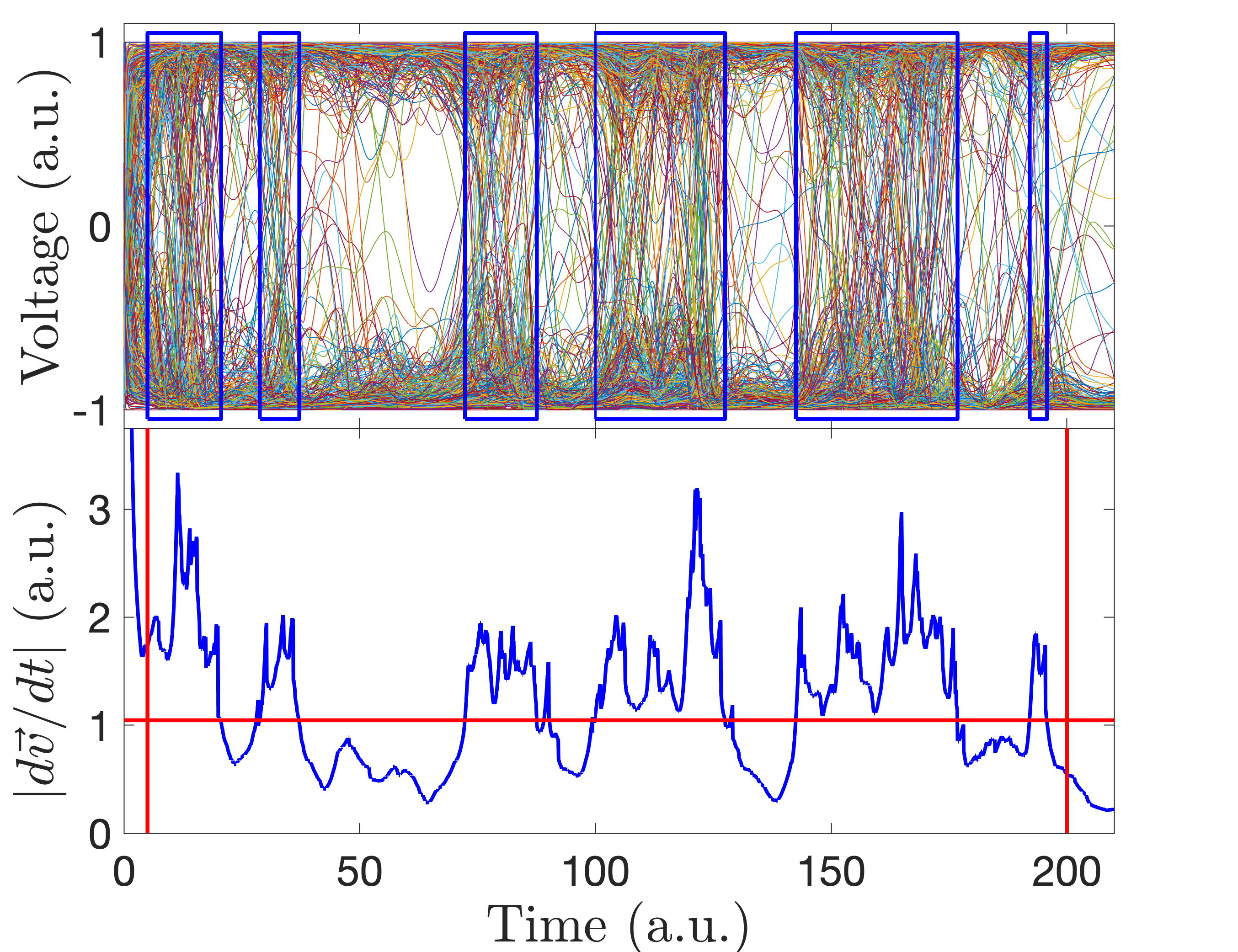}
	\caption{Top panel: Voltages for the solution of a 420-variable random 3-SAT problem. The blue boxed regions correspond to avalanche activity. Bottom panel: The magnitude of the vector of all voltage time derivatives. An initial short transient (finishing at the leftmost vertical red line) and the very final approach to solution (starting at the rightmost vertical red line) are ignored in the calculation of avalanches. The avalanches are identified by a threshold, determined by adding 25\% of the minimum-to-maximum distance to the lowest voltage derivative value (horizontal red line). The open regions in the top panel correspond to regions absent of collective threshold crossings of the voltage derivative.}
	\label{fig:thresholding}
\end{figure}

With these preliminaries we can now discuss the transient dynamics of DMMs and argue that the size of the generated 
avalanches (of the voltages at the gate terminals) must follow a probability distribution typical of critical branching processes. We first note 
that at the initial time of the dynamics a general DMM finds itself in an unstable (unsatisfied) state. The voltages at the different terminals of the gates then start evolving, and at some intervals of time some of them undergo sudden transitions, thus creating avalanches (see typical voltage trajectories in Fig.~\ref{fig:thresholding}, top panel)~\cite{topo}. Additionally, the memory variables in DMMs have much slower dynamics than the voltage variables~\cite{DMM2,DMMperspective}. This implies that each avalanche is independent of the others generated (mean-field condition). 


Now, every time a given voltage flips from +1 to $-1$, or vice-versa, so that its corresponding Boolean variable changes from logical 1 to logical 0, or the reverse, on average, it will only have enough strength (power) to affect one other voltage in the circuit (its ``offspring''). In turn, this ``offspring'' voltage, on average, will have enough strength to only affect at most one other voltage at the next time step, and so on. 

Since all voltages in the circuit are equally important, the distribution of the number of voltages affected by a given voltage must be the same for each individual voltage at every time step (a ``generation'' step), and independent of both the number of flipped voltages at that time step and the number of affected voltages (offspring). Therefore, the flipping of a single voltage gives rise to a Poisson-distributed process where the average number of affected variables is $\mu\rightarrow 1$. 

Under these conditions, the number of ``descendants'' of a flipped voltage (the size of the avalanches) is an integer random variable, $S$, described by the Borel distribution $p_S = (\mu S)^{S-1}e^{-\mu S}/S!$ \cite{borel_derivation}. The expectation value of $S$ is given by $\langle S \rangle=1/(1-\mu)$. Therefore, due to $\mu\rightarrow 1$, DMMs must showcase a critical branching process. In fact, in the limit of $\mu\rightarrow 1$, the Stirling approximation of the Borel distribution is proportional to $S^{-3/2}$, 
namely a scale-free distribution~\cite{borelApprox1, borelApprox2}. 

\section{DMM equations of motion}

We can now corroborate this theoretical analysis with actual numerical results. We first design a DMM for solving 3-SAT problems. 
Since the choice of the dynamical system representing a DMM is not unique~\cite{DMM2,DMMperspective}, we choose one very similar to the one employed in Ref.~\cite{spinglass} to find the ground state of the Ising spin-glasses. 

In a 3-SAT problem, clauses take the form $(\ell_i\ \ell_j\ \ell_k)$, where $\ell_i$ is a literal associated with a Boolean propositional variable $y_i$, and can be either $\ell_i=y_i$ or $\ell_i=\lnot y_i$.
The variables, $y_i$, are transformed to continuous variables, $v_i$, representing terminal voltages on the self-organizing OR gates (see Fig. \ref{fig:circuit}) \cite{DMM2,Bearden}. The voltages are bounded, $v_i \in [-1,1]$, with $v_i \geq 0 $ transformed to $y_i=1$ and $v_i < 0$ transformed to $y_i=0$. The negation operation used by literals is trivially performed on the voltages by multiplying them by $-1$.

We then convert the $n$-th clause to a dynamical clause by interpreting literals as voltages,
\begin{equation}
C_n(v_{i},v_{j},v_{k}) = \frac{1}{2}\textrm{min}(1 \pm v_{i},1 \pm v_{j},1 \pm v_{k}),
\label{eq:clause}
\end{equation}
where subtraction is used if $\ell_i=y_i$ and addition is used if $\ell_i=\lnot y_i$, with $C_n \in [0,1]$. When the clause is satisfied we have $C_n = 0$. 

The dynamics of the voltages are influenced by the dynamical clauses in which the voltages appear~\cite{spinglass},

\begin{equation}
\begin{split}
\frac{d}{dt}v_i = -\sum_{n} x_{s,n}x_{f,n}G_{n,i}(v_i,v_j,v_k)\ +\\
(1-x_{f,n})R_{n,i}(v_i,v_j,v_k),
\end{split}
\label{eq:voltages}
\end{equation}
where the sum is taken over all clauses, $C_n$, in which $v_i$ is present.  The initial condition for the voltages is chosen randomly in the interval $[-1,1]$, and the solution is found when $C_n=0$ for all clauses. 
The memory variables, $x_{s,n}$ and $x_{f,n}$, along with the functions $G_{n,i}$ and $R_{n,i}$, are discussed below.

Each clause has its own memory variables, $x_{f,n}$ and $x_{s,n}$, containing information of past dynamics. The ``fast'' memory variable, $x_{f,n}$, determines the state of satisfiability of the clause in the recent past. Its dynamics are governed by
\begin{equation}
\frac{d}{dt}x_{f,n} = \beta (x_{f,n}(1-x_{f,n}))^{1/2}(C_n(v_{i},v_{j},v_{k})-\gamma),
\label{eq:xfast}
\end{equation}
where we have chosen $\beta = 1/2$ and $\gamma = 1/8$. The fast memory variable is bounded, $x_{f,n} \in [0+\epsilon,1-\epsilon]$, with the offset, $\epsilon = 10^{-3}$, such that $x_{f,n} = 0+\epsilon$ is interpreted to mean the clause has been satisfied for a period of time in the recent past, and $x_{f,n} = 1-\epsilon$ means the clause has been unsatisfied for a period of time in the recent past. The offset is used to remove spurious steady-state solutions from Eq. (\ref{eq:xfast}).

The role $x_{f,n}$ has on Eq.~(\ref{eq:voltages}) is to switch between the first and second terms in the summation.   It can be seen that $x_{f,n}$ continuously switches between two modes: search for a satisfying assignment and hold the satisfying assignment. The first term in the summation contains a ``gradient-like'' function, $G_{n,i}(v_i,v_j,v_k)$, that tries to satisfy clause $n$ by changing $v_i$,
\begin{equation}
G_{n,i}(v_{i},v_{j},v_{k}) = \frac{ \partial}{ \partial v_i} (1 \pm v_i)\textrm{min}[(1 \pm v_j),(1 \pm v_k)],
\label{eq:grad}
\end{equation}
where the sign is chosen as in Eq.~(\ref{eq:clause}).

The second term in the summation of Eq.~(\ref{eq:voltages}) contains a ``rigidity'' function~\cite{spinglass}, $R_{n,i}(v_i,v_j,v_k)$, which either tries to pull $v_i$ towards an assignment ($v_i=\pm1$) that makes $C_n=0$ if $v_i$ is the voltage closest to satisfying the clause, or does nothing to $v_i$, if $v_j$ or $v_k$ is the voltage closest to satisfying the clause, namely
\begin{equation}
\begin{split}
R_{n,i}(v_{i},v_{j},v_{k}) &= \\
&\begin{cases} 
v_i \pm 1, & C_n(v_{i},v_{j},v_{k}) = \frac{1}{2}(1\pm v_i)  \\
0, & \textrm{otherwise}
\end{cases}
\end{split}
\label{eq:rigid}
\end{equation}
Again, the signs are chosen as in Eq.~(\ref{eq:clause}).

The slow memory variable, $x_{s,n}$, adds weight to the gradient-like term for clause $n$ in Eq.~(\ref{eq:voltages}), but does not affect the rigidity term. The additional weight promotes the ability to overcome the rigidity terms associated with other clauses. In essence, $x_{s,n}$ acts like a memory-assisted current generator that injects current into the circuit to push the DMM towards the solution, as originally conceived in Ref.~\cite{DMM2}. This variable is also bounded, $x_{s,n} \in [1, M]$, where $M$ is the number of variables in the problem\footnote{In practice, it is not necessary to have such a large upper bound.}, 
and we choose as its dynamics: 
\begin{equation}
\frac{d}{dt}x_{s,n} = \alpha x_{s,n}(x_{f,n} - C_n(v_{i},v_{j},v_{k})),
\label{eq:xslow}
\end{equation}
where we have chosen $\alpha = 1/100$. The slow memory variable will grow while its associated clause is unsatisfied, thereby giving the literals within that clause added weight to influence the dynamics of the voltages. In effect, $x_{s,n}$ contains memory of how often the clause was unsatisfied while traversing the phase space. We choose to initialize both the slow and fast memory variables as 1 for all clauses. 


\section{Results}

With these DMM equations we have solved benchmark problems from previous SAT competitions. The random 3-SAT benchmarks were taken from www.satcompetition.org, and correspond to a ratio between clauses and variables of $4.3$. We have solved instances of random 3-SAT for variable sizes 420, 460, 500, 540, and 600. For each variable size we have extracted data from 100 solutions found from random initial conditions. Once a solution is found, we analyze the data for avalanches.

To identify an avalanche we analyze the magnitude of the time derivative of the vector of all the voltages, $|d\vec{v}(t)/dt|$, which we refer to as the ``voltage derivative''. The random initial conditions cause a large spike in $|d\vec{v}(t)/dt|$, as seen in the bottom panel of Fig. \ref{fig:thresholding}. This short transient is then ignored by removing the first 5 units of time from the voltage derivative (leftmost red vertical line). When the DMM has found a solution, the voltage derivative will approach zero. We then choose to ignore also the last 10 units of time of the voltage derivative (rightmost red vertical line), so the minimum of the voltage derivative is not found at the time boundary of the simulation\footnote{Occasionally, the interval of 10 units of time is not large enough. When this occurs, the cutoff is incremented by 1 unit of time, and the minimum is checked again. The process is repeated until the minimum is not found on the time boundary.}.

We show in the top panel of Fig.~\ref{fig:thresholding} all voltages in the solution of a random 3-SAT problem. From the figure it is easy to identify ``open'' regions that are best described by a lack of collective flipping of voltages. These regions are characteristic of an absence of avalanche events. The identified avalanche events are enclosed in blue boxes.

\begin{figure}[t]
	\centering
	\includegraphics[width=8.5cm]{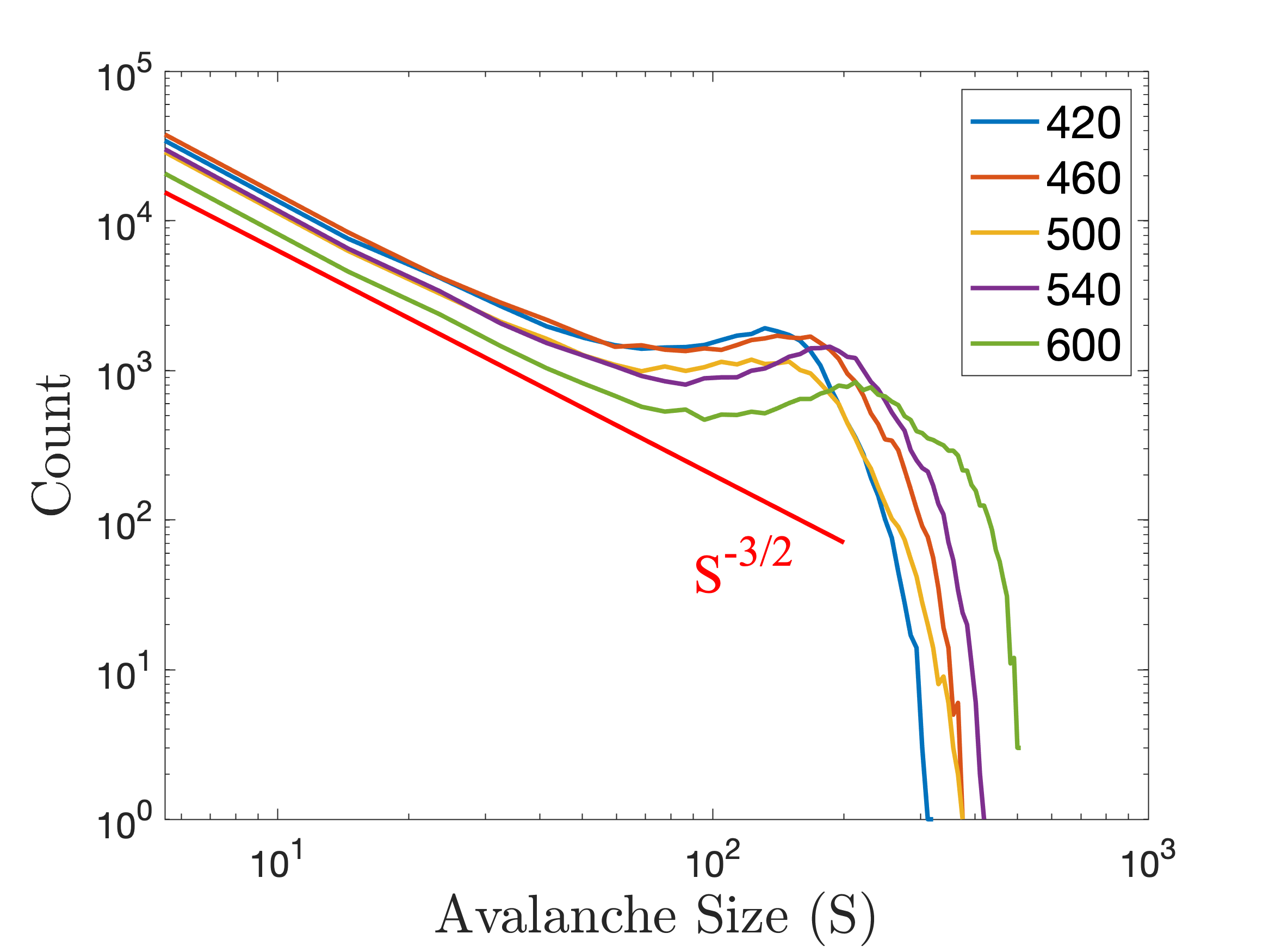}
	\vspace{-0.1cm}
	\caption{Distribution of avalanche sizes, $S$, for different variable sizes of random 3-SAT instances. The red line is proportional to $S^{-3/2}$ (a critical Borel distribution), and is shown for comparison. The curve for each variable size originates from 100 solutions found from Eqs.~(\ref{eq:voltages}),~(\ref{eq:xfast}), and~(\ref{eq:xslow}), with Eq. (\ref{eq:voltages}) having random initial conditions.}
	\label{fig:distribution}
\end{figure}

Once the short initial transient and the very final approach to solution are eliminated from the voltage derivative, we find its maximum and minimum. A threshold is calculated by adding 25\% of the distance between minimum and maximum to the minimum of the voltage derivative. When the voltage derivative rises above the threshold we assume that an avalanche begins, and when the voltage derivative drops below the threshold the avalanche ends. Between these two events, we check how many voltages change sign. The size of the avalanche, $S$, is defined as the number of voltages that change sign between threshold crossings. 

Our scheme for determining avalanches is, of course, not without uncertainty. For instance, it is possible for multiple avalanches to be identified as a single avalanche when they are close together in time, and thresholding may also miss the voltages that flipped immediately before and after an avalanche event.

To account for the uncertainty introduced by thresholding we choose to bin the data of our numerical simulations. The bin size, $R$, is chosen using the Freedman-Diaconis rule~\cite{FDRule}, $R = 2IQR(S)A^{-1/3}$, rounded to the nearest integer, where $IQR(S)$ is the interquartile range of the distribution of avalanche sizes, $S$, and $A$ is the number of avalanches observed. For the 600-variable instances, the Freedman-Diaconis rule rounds to $R=9$. For better comparison, we have applied this bin size to all variable sizes.

The results of the analysis are shown in Fig.~\ref{fig:distribution}. Using a power-law fit we find that the initial portion of the distribution is proportional to $S^{-\tau}$, with $\tau = 1.51\pm 0.02$, which is in very good agreement with the one predicted by our mean-field theory, thus giving support to the 
hypotheses made in that analysis.

Note that the distribution in Fig. \ref{fig:distribution} has the characteristic ``bump'' seen in many finite-size power-law distributions \cite{pruessner_2012}. We attribute the bump to mis-identifications made by the thresholding process. For example, in Fig. \ref{fig:thresholding} we see for $100<t<125$ the thresholding has identified one avalanche. If the thresholding were to be raised slightly, the thresholding would identify three avalanches. However, we have found that changing the thresholding percentage has a negligible effect on the scale-free trend before the bump, because the composite avalanches are comprised of smaller avalanches, and thus different thresholds simply correspond to different sampling from the same distribution.

\section{Conclusions}

In conclusion, we have provided analytical arguments, supported by numerical results, that memcomputing machines (machines that use memory to process information) undergo a critical branching process with exponent $3/2$ during their transient dynamics. The dynamics of DMMs then 
share some of the same features observed in many non-equilibrium phenomena encountered in Nature, and demonstrate the 
rich phenomenology these dynamical systems showcase, which is behind their ability to solve complex problems efficiently. 



%
%
%
%

{\it Acknowledgments --} This material is based upon work supported by the National Science Foundation Graduate Research Fellowship under Grant No. DGE-1650112. S.B. acknowledges support from the Alfred P. Sloan Foundation's Minority Ph.D. Program. M.D. acknowledges partial support from the Center for Memory and Recording Research at UC San Diego.

\end{document}